\documentstyle[12pt,aaspp4]{article}
\begin{document}
\parindent=1.0cm

\title{The Metallicity of the Red Giant Branch in the Disk of NGC 6822}

\author{T. J. Davidge \footnote[1]{Visiting Astronomer, Canada-France-Hawaii
Telescope, which is operated by the National Research Council of Canada, 
the Centre National de la Recherche Scientifique, and the University of 
Hawaii}}

\affil{Canadian Gemini Office, Herzberg Institute of Astrophysics,
\\National Research Council of Canada, 5071 West Saanich Road, 
\\Victoria, B.C. Canada V9E 2E7\\ {\it email: tim.davidge@nrc-cnrc.gc.ca}}

\begin{abstract}

	Deep $J, H,$ and $K'$ images obtained with the Canada-France-Hawaii 
Telescope adaptive optics system are used to investigate the metallicity of red 
giant branch (RGB) stars in three fields in the disk of the Local Group dwarf 
irregular galaxy NGC 6822. The slope of the RGB on the $(K, J-K)$ 
color-magnitude diagrams indicates that $\overline{[Fe/H]} = -1.0 \pm 0.3$. The 
locus of the RGB is bluer than that of globular clusters with the same RGB 
slope, by an amount that is consistent with the 
majority of RGB stars in these fields having an age 
near 3 Gyr. It is demonstrated that if RGB stars in NGC 6822 are this young 
then the metallicity computed from the RGB slope may be $\sim 0.05$ dex 
too low. Finally, the metallicity computed from the RGB slope is 
lower than spectroscopic-based metallicities for young 
stars in NGC 6822, and it is concluded that $\frac{\Delta [Fe/H]}{\Delta t} = 
-0.2 \pm 0.1$ dex Gyr$^{-1}$ in NGC 6822 during the past few Gyr.

\end{abstract}


\section{INTRODUCTION}

	NGC 6822 is an isolated dwarf irregular (dIrr) galaxy belonging 
to the `Local Group Cloud' (Mateo 1998). As the nearest isolated dIrr galaxy, 
NGC 6822 is an important laboratory for probing 
the evolution of this type of system. It is therefore 
not surprising that the stellar content of NGC 6822 has been the subject of a 
number of investigations, and these have found stars spanning a broad range 
of ages. 

	The youngest stars in NGC 6822 are distributed in an irregular 
fashion in and around the central bar (Hodge 1980; Hodge et al. 
1991; Marconi et al. 1995; Gallert et al. 1996a; Wyder 2001). Gallert et al. 
(1996b) conclude that the star formation rate (SFR) in NGC 6822 has been 
constant over much of the age of the galaxy, with the exception of a 
drop in activity during intermediate epochs. However, the SFR 
has increased during the past one Gyr (Marconi 
et al. 1995; Gallert et al. 1996a; Wyder 2001), and there are morphological 
signatures that are suggestive of a tidal interaction (Hutchings, Cavanagh, \& 
Bianchi 1999; de Blok \& Walter 2000), which may have triggered the recent 
increase in star-forming activity. NGC 6822 
is also surrounded by an extended stellar halo that has been traced with 
Carbon stars (Letarte et al. 2002), suggesting that during intermediate epochs 
star formation was distributed over a larger area than is currently the case.

	Tolstoy et al. (2001) conclude that [Fe/H] in NGC 6822 grew steadily 
with time from early epochs until 3 Gyr in the past, at which point the pace of 
chemical enrichment accelerated due to an increase in the SFR. For 
comparison, Wyder (2001) suggests that NGC 6822 may have experienced a rapid 
initial chemical enrichment, and that [Fe/H] has been constant or only slowly 
changing with time during the past 7 Gyr with [Fe/H] $= -1$. However, Wyder 
(2001) also emphasizes that the age-metallicity degeneracy that plagues 
photometric studies of old populations renders his chemical enrichment history 
of NGC 6822 very uncertain more than 3 Gyr in the past. 

	There is a growing body of spectroscopic observations of sources in 
NGC 6822 that can be used to investigate directly the chemical enrichment 
history of the galaxy. Pagel, Edmunds, \& Smith (1980) and Skillman, 
Terlevich, \& Melnick (1989) investigated the spectra of HII regions in NGC 
6822, and found [O/H] that is intermediate between the SMC and LMC. The 
abundances obtained from HII regions are consistent with those measured from 
the absorption line spectra of bright supergiants (Muschielok et al. 
1999; Venn et al. 2001). Chandar, Bianchi, \& Ford (2000) 
discussed spectroscopic observations of HII regions outside of the central 
regions of NGC 6822, and found [O/H] that is significantly different 
from that measured by Pagel et al. (1980) and Skillman et al. (1989), 
suggesting that [O/H] may vary across the galaxy. Venn et al. (2001) noted that 
an abundance gradient is consistent with the highest 
quality abundance measurements in NGC 6822, but also 
stressed that additional data are needed to characterize with confidence 
the radial chemical properties of the galaxy. Chandar et al. (2000) also 
investigated intermediate age clusters in NGC 6822, and found metallicities 
that are $0.5 - 1.0$ dex lower than in clusters with comparable ages in the 
LMC and SMC. Tolstoy et al. (2001) measured the 
strength of the near-infrared Ca II lines in a sample of RGB stars, and found 
[Fe/H] ranging from --2 to --0.3, with a peak near [Fe/H] $= -0.9$. The 
metallicity distribution function (MDF) constructed by Tolstoy et al. (2001) 
is skewed by a few tenths of a dex to lower values than the MDF of LMC giants 
constructed by Cole, Smecker-Hane, \& Gallagher (2000), which peaks at [Fe/H] = 
$-0.57$, with $\overline{[Fe/H]} = -0.64$. Galaxies follow a relation between 
mass and metallicity that is thought to be a consequence of feedback 
following periods of intense star formation, and 
the relation predicted by Dekel \& Silk (1986) 
predicts that the metallicities of the LMC and NGC 6822 should differ by $\sim 
0.2$ dex, which is consistent with the observed metallicities of these systems.

	In the present study, deep $JHK$ observations of three fields in the 
disk of NGC 6822 are used to measure the metallicity of red giants, and 
thereby provide additional insight into the chemical enrichment 
history of this region of the galaxy during old and intermediate epochs. 
The metallicity is estimated from the slope of the RGB on the $(K, J-K)$ 
CMD. The slope of the RGB in infrared CMDs is insensitive to uncertainties 
in the reddening and the photometric calibration, and has been calibrated as a 
metallicity indicator using CMDs of globular clusters spanning a range 
of metallicities (e.g. Ferraro et al. 2000; hereafter F2000). The RGB slope is 
also insensitive to the age-metallicity degeneracy that complicate efforts to 
estimate metallicities from the color of the RGB.

	NGC 6822 is viewed at a low Galactic latitude, and so there is 
significant extinction and contamination from foreground stars. The Schlegel, 
Finkbeiner, \& Davis (1998) dust maps give a foreground reddening of $E(B-V) = 
0.24$ near the center of NGC 6822, and this is adopted as the baseline for 
the present study, although there is significant internal extinction 
near star forming regions in NGC 6822 (\S 5). A distance 
modulus of 23.49, which was computed from Cepheids and is consistent with the 
brightness of the RGB-tip (Gallart, Aparicio, \& Vilchez 1996), 
is also adopted here. 

	The paper is structured as follows. The observations and reductions 
are described in \S 2, while the CMDs, luminosity functions 
(LF)s, and two-color diagrams (TCDs), of the fields are 
discussed in \S 3. The metallicity of the RGB is estimated from the slope 
of the RGB on the $(K, J-K)$ CMD in \S 4. A summary and discussion 
of the results follows in \S 5.

\section{OBSERVATIONS \& REDUCTIONS}

	The data were recorded during the nights of September 9/10 and 10/11 
2000 (UT) with the Canada-France-Hawaii Telescope (CFHT) Adapative Optics 
Bonnette (AOB) and KIR camera. A detailed description of the AOB, 
which uses natural guide stars to monitor wavefront distortions, has been 
given by Rigaut et al. (1998). KIR is the dedicated 
infrared imager for the AOB and contains a $1024 
\times 1024$ Hg:Cd:Te array, with each pixel subtending 0.034 arcsec on a side.

	Three fields in the disk (i.e. outside of the bar) of NGC 6822 
were imaged through $J, H,$ and $K'$ filters. Each field contains 
a moderately bright ($R \leq 15$) AO guide star, and the names and co-ordinates 
of these stars are listed in Table 1. The field locations 
are marked in Figure 1, which shows a portion of the Digital Sky Survey 
centered on NGC 6822.

	Each AO guide star was positioned near the center of the KIR science 
field to reduce the effects of anisoplanicity near the edges of the detector. 
Five 60 sec exposures were recorded in each filter at the corners of a 
$0.5 \times 0.5$ arcsec square dither pattern, and the total exposure time 
per filter is thus 60 sec $\times$ 5 exposures $times$ 4 dither positions 
= 1200 secs. The sky was photometric when these data were 
recorded, and the seeing was very good. Stars in the corrected images have FWHM 
= $0.10 - 0.15$ arcsec, and the angular resolution of the $K'$ images is the 
telescope diffraction limit.

	The processing sequence consisted of the following steps: (1) dark 
subtraction, (2) division by a dome flat, which was constructed by differencing 
images of a dome spot taken with the dome lights on and off to remove thermal 
artifacts, (3) the subtraction of a calibration image to remove interference 
fringes and the thermal signatures of warm objects along the light path, which 
was constructed by combining images of different fields to suppress 
sky sources, and (4) the subtraction of the DC sky level, 
which was measured separately for each exposure. The results 
were aligned on a field-by-field basis to correct for the dither offsets, 
and then median-combined to reject cosmic rays and bad 
pixels. The combined images were trimmed to the 
region having a full 20 minute exposure time, and the final $K'$ images 
are shown in Figures 2, 3, and 4. The AO guide star 
is the bright central source in each figure, and the first diffraction ring 
can be seen around moderately bright stars in each of the fields.

\section{RESULTS}

\subsection{Photometric Measurements}

	Stellar brightnesses were measured with the point 
spread function (PSF) fitting routine ALLSTAR (Stetson \& Harris 1988), which is
part of the DAOPHOT (Stetson 1987) photometry package. A single PSF was 
constructed for each field $+$ filter combination. 
While anisoplanicity causes the PSF to change with angular distance from the 
AO guide star, observations of globular clusters with the CFHT AOB indicate 
that, at least for moderate levels of AO compensation during 
typical Mauna Kea seeing conditions, the PSF typically 
changes on a characteristic angular scale that exceeds the KIR field 
(e.g. Davidge \& Courteau 1999; Davidge 2001). The relatively large isoplanatic 
patch is a consequence of the superb seeing conditions on Mauna Kea, and the 
PSF will change over smaller angular scales at sites with poorer 
atmospheric conditions. 

	There are three pieces of evidence that justify the use of a 
single PSF. First, a classical signature of anisoplanicity is the 
radial elongation of stellar images near the field edges (McClure et al. 1991), 
with the major axis pointing towards the guide star, and this distortion 
is not seen in these data. Second, tight, well-defined CMDs are produced from 
these data when a uniform PSF is assumed (\S 3.2), demonstrating that 
aniosplanicity causes only moderate scatter in the photometric 
measurements. Third, the encircled energy curves of PSFs constructed 
from stars in two different radial intervals from the AO guide star in the $K'$ 
Field 1 data, which are shown in Figure 5, demonstrate the radial 
stability of the PSF. The radial intervals used to construct the 
encircled energy curves sample comparable areas in the KIR science field, and 
the PSFs were constructed from similar numbers of stars. The two curves in 
Figure 5 agree to within a few percent inside the PSF-fitting radius.

	The photometric calibration is based on observations of UKIRT 
faint standard stars (Hawarden et al. 2001) that were recorded throughout the 2 
night observing run. The wings of AO-corrected PSFs extend out to 
distances in excess of an arcsec, and so the brightnesses of the standard stars 
were measured in 2 arcsec radius apertures to sample as much light as possible; 
measurements with larger apertures were found not to contain additional signal. 
The uncertainty in the photometric zeropoints, computed from the residuals 
about the mean, is $\pm 0.04$ mag in $K'$. For consistency with the standard 
star observations, 2 arcsec radius aperture measurements of the PSF stars 
were used to calibrate the NGC 6822 observations. These aperture measurements 
were made after all stars but those used to construct the PSF were subtracted 
from the images. The use of large radius aperture measurements to calibrate the 
data greatly reduces the sensitivity to field-to-field variations in the
Strehl ratio (e.g. Davidge 2002). 

	Artificial star experiments, in which scaled versions of the PSFs were 
added to the processed images using the DAOPHOT ADDSTAR routine, were run 
to estimate completeness and assess the scatter due to photometric errors. The 
artificial stars were assigned brightnesses and colors falling along the RGB 
ridgeline. A fixed PSF was assumed for each field, and so these simulations do 
not take into account anisoplanicity. The brightnesses of the artificial stars 
were measured using the same procedure applied to real stars in NGC 6822. These 
experiments indicate that incompleteness sets in at $K = 20$ in Field 1, and $K 
= 20.5$ in Fields 2 and 3, while the 50\% completeness levels occur near 
$K = 20.5$ in Field 1, and $K = 21.0$ in Fields 2 and 3.

\subsection{$(K, H-K)$ and $(K, J-K)$ CMDs}

	When compared with $J-H$ or $J-K$, the $H-K$ color changes slowly with 
effective temperature for all but the coolest stars, and so the width of the 
RGB in the $(K, H-K)$ CMD is used here as a means of constraining the 
uncertainties in the photometric measurements due to anisoplanicity. 
The $(K, H-K)$ CMDs of the NGC 6822 fields are shown in Figure 6, and 
the RGB is clearly evident as a vertical sequence near $H-K = 0$. 
The scatter envelope predicted from the artificial star experiments is slightly 
smaller than the observed width of the RGB. In particular, the standard 
deviation in the $H-K$ colors of stars with $K$ between 18 and 19 
in each field is typically $\pm 0.06$ mag, while the dispersion 
predicted from the artificial star experiments in this same magnitude 
interval is $\pm 0.04$ mag. The residual scatter due 
to the intrinsic properties of stars in NGC 6822, differential reddening, 
and anisoplanaticism, which can be computed by subtracting the predicted 
scatter from the measured scatter in quadrature, is then $\pm 0.04$ mag. 
This indicates that anisoplanicity does not introduce scatter in the $H-K$ 
colors at a level in excess of a few hundredths of a mag, and this is 
of the same order as the uncertainties predicted from the 
encircled energy curves that are compared in Figure 5. 

	The $(K, J-K)$ CMDs of the three fields are compared in Figure 7. 
The RGB is the dominant feature in each CMD, indicating that these fields 
are dominated by old or intermediate age populations. 
While the upper RGBs of Fields 1 and 2 
are not well defined, the RGB of Field 3 is curved near the 
bright end, indicating that RGB stars in this region of NGC 6822 are not 
extremely metal-poor. The onset of the RGB in all three fields occurs near $K = 
17$. Finally, given the low Galactic latitude of NGC 6822, it is likely that 
some of the stars on the $(K, J-K)$ CMDs that do not fall along the RGB 
belong to the foreground, especially those that are redder than the RGB. 
Some of the objects in Fields 1 and 3 with $J-K \sim 0$ 
may be bright main sequence stars in NGC 6822. 

	The photometric errors are largest, and hence most likely to dominate 
over intrinsic star-to-star differences in photometric properties, 
at the faint end, and there is good agreement between the predicted and 
observed scatter in the $(K, J-K)$ CMDs when $K \geq 19$. 
However, at the bright end the RGB on the $(K, J-K)$ CMD is 
noticeably wider than expected from photometric errors alone. 
A number of factors could introduce scatter in the CMD, 
and these include (1) the AGB, (2) differential reddening, (3) a range 
in metallicity, and (4) an age dispersion. Limits on age and metallicity 
dispersions in the disk of NGC 6822 are calculated in \S 4.

\subsection{$K$ LFs and $(J-H, H-K)$ TCDs}

	The $K$ LFs of stars that are detected in both $H$ 
and $K'$ and have $H-K$ colors within $\pm 0.2$ mag of the RGB ridgeline on the 
$(K, H-K)$ CMD are shown in Figure 8. A composite LF was also created 
by summing the LFs of the individual fields to reduce the small 
number statistics that dominate the LFs of the individual fields near the 
RGB-tip, and the result is shown in the bottom panel of Figure 
8. The LFs follow power laws at the faint end, and the 
method of least squares was used to fit a power-law to the entries with $K$ 
between 18 and 20 in the summed LF; this brightness interval was 
selected to fit the power law since it is fainter than the RGB-tip, and 
incompleteness is not an issue. The fitted power law has an exponent $0.23 \pm 
0.06$, which agrees with the exponents measured in Galactic globular clusters 
by Davidge (2000, 2001).

	The $(J-H, H-K)$ TCD can be used to identify objects with non-stellar 
near-infrared spectral-energy distributions (SEDs), such as emission line 
galaxies, as well as stars with extreme SEDs, such as long period variables 
(LPVs). The TCDs of the three fields, corrected for foreground extinction, are 
shown in Figure 9. The majority of objects fall along the globular cluster 
giant branch sequence, which is not unexpected given that the RGB is the 
dominant feature in the NGC 6822 CMDs. There are also a modest number of stars 
that appear to be LPVs, as well as some sources with non-stellar SEDs.

\section{THE RGB SLOPE AND METALLICITY}

	Small number statistics complicate efforts to track the upper portions 
of the RGB on CMDs, and this affects efforts to measure the slope of 
the RGB. Therefore, the data from all three fields were combined 
to better define the upper portions of the RGB. 
This was done by calculating normal points for each field, and then 
computing a mean normal point sequence for all three fields. 
Normal points were calculated by finding the 
mean $J-K$ color in $\pm 0.25$ mag bins along the $K$ axis of each $(K, J-K)$ 
CMD, with an iterative $2.5 \sigma$ rejection scheme to suppress outliers. 
A single bulk shift along the $J-K$ axis was then applied to all of the 
datapoints in each field such that the normal point sequence for that field 
best matched the mean normal point sequence. The field-to-field scatter 
among the normal point sequences is $\pm 0.05$ mag, which is comparable to the 
uncertainty in the $J-K$ calibration based on the standard star measurements. 

	The composite (M$_{K}, (J-K)_0$) CMD, assuming the reddening and 
distance modulus discussed in \S 1, is shown in the left hand panel of Figure 
10. The data from the various fields are well mixed with no systematic 
differences near the RGB-tip, which is consistent with the three fields 
having similar RGB morphologies. The upper RGB is better defined than was 
the case from the CMDs of the individual fields, and M$_{K}^{RGBT} \sim -6.5$. 

	To suppress the effects of non-RGB stars and photometric scatter on 
the slope measurement, a final set of normal points was computed from the 
composite CMD, and the sequence defined by these normal points is 
shown in the middle panel of Figure 10. 
The slope measured from a least squares fit to the normal 
points is $\frac{\Delta (J-K)}{\Delta K} = -0.108 \pm 0.007$. This slope 
measurement serves as the basis for computing [Fe/H] in the disk of NGC 6822. 

	F2000 give a relation between RGB slope and metallicity in their 
Figure 12a that was determined from the CMDs of globular clusters 
spanning a range of metallicities; however, F2000 measured RGB slopes 
using individual CMD datapoints, rather than 
normal points. This is an important consideration as the use of normal points 
computed at uniform magnitude intervals effectively assigns each portion of the 
RGB equal weight when measuring the RGB slope, whereas less weight is 
assigned to the uppermost portion of the RGB when individual CMD 
points are used, since the number of stars per unit brightness interval 
decreases towards the RGB-tip.

	Table 2 of F2000 lists the colors of various 
globular cluster RGBs at several M$_K$, and these were used to 
calibrate the metallicity dependence of RGB slopes measured from normal points. 
The slopes of the clusters in the F2000 sample were measured 
from the entries in their Table 2, supplemented with M$_{K}^{RGBT}$ from their 
Table 6 and RGB-tip colors from the CMDs plotted in their Figure 1. NGC 6528 
and NGC 6553 were not considered as there is significant scatter on the upper 
RGBs of these clusters in Figure 1 of F2000. A linear least 
squares fit to these slope measurements gives:

\noindent\hspace*{3.0cm}{[Fe/H] = $(-16.95 \pm 2.28) \times \frac{\Delta J-K}{\Delta K} - (2.87 \pm 0.07)$ \hspace*{1.0cm} (1)}

	This normal point-based calibration differs from that determined from 
measurements of individual stars. For example, whereas a system with [Fe/H] 
$= -0.7$ is predicted to have $\frac{\Delta J-K}{\Delta K} = -0.097$ from the 
relation in Figure 12a of F2000, Equation 1 predicts 
that $\frac{\Delta J-K}{\Delta K} = -0.128$ for this same system.

	Equation 1 indicates that [Fe/H] $= -1.0 \pm 0.3$ for the disk of NGC 
6822, where the quoted error includes the uncertainties in the coefficients 
in Equation 1 and the uncertainty in the slope measurement. This metallicity 
can be checked using the RGB-tip brightness in $K$ and the color of the RGB, 
both of which are sensitive to metallicity.

	M$_{K}^{RGBT}$ becomes brighter as [Fe/H] increases. Models 
by Girardi et al. (2000, hereafter G2000) predict that 
M$_K^{RGBT} = -5.6$ in an old population with [Fe/H] $= -1.6$, and M$_K^{RGBT} 
= -6.3$ when [Fe/H] $= -0.6$. The empirical calibration of F2000 
gives slightly brighter RGB-tip values, with M$_K^{RGBT} = -6.0$ and 
M$_K^{RGBT} = -6.6$ for globular clusters with [Fe/H] $= -1.6$ and $-0.7$. 
The G2000 models also predict that at a fixed metallicity M$_K^{RGBT}$ becomes 
fainter as age decreases, although the differences in M$_K^{RGBT}$ due to age 
are relatively modest compared with those due to metallicity; for example, 
between t = 16 Gyr and 3 Gyr the z=0.001 (i.e. [Fe/H] $= -1.2$) G2000 models 
predict that M$_K^{RGBT}$ dims by only 0.2 mag. The F2000 calibration predicts 
that M$_{K}^{RGBT} = -6.4 \pm 0.2$ for an old system with [Fe/H] $= -1.0 \pm 
0.3$, which agrees with the RGB-tip brightness measured for the disk of 
NGC 6822.

	While the $K$ brightness of the RGB-tip in the disk of NGC 6822 is 
consistent with the measured metallicity, the RGB 
ridgeline is bluer than expected for an old, 
[Fe/H] $= -1$ population. This is demonstrated in the middle panel of Figure 
10, where the RGB sequences of 47 Tuc ([Fe/H] $= -0.7$), NGC 6121 
([Fe/H] $= -1.2$), and NGC 7078 ([Fe/H] $=-2.1$) from F2000 are compared 
with the NGC 6822 normal point sequence. The NGC 6822 giant branch 
falls blueward of both the 47 Tuc and NGC 6121 sequences, even though 
the metallicity estimated from the RGB slope falls midway between the 
metallicities of these clusters. This color offset may indicate that 
RGB stars in NGC 6822 and globular clusters have different ages, in the sense 
that stars in NGC 6822 are younger. 

	The age sensitivity of RGB sequences is investigated in the right hand 
panel of Figure 10, which shows z = 0.001 isochrones of G2000 for populations 
with log(t$_{Gyr}$) = 9.3, 9.6, 9.9, and 10.2. The NGC 6822 sequence in 
Figure 10 falls $\sim 0.1$ mag blueward of the expected location for an 
old stellar system having the same metallicity, and this is consistent with 
the RGB stars in NGC 6822 having a typical age of $\sim 3$ 
Gyr if globular clusters have ages $\geq 10$ Gyr. Such a relatively 
young age for RGB stars in NGC 6822 is consistent with a star-forming 
history that is not weighted towards early epochs, but in which there has been 
significant recent star formation.

	An age difference between NGC 6822 and globular clusters will have 
only a modest impact on the metallicity estimated from the RGB slope, and the 
G2000 isochrones in the right hand panel of Figure 10 can be used to 
quantify any age-related systematic effects. 
To simulate the process used to create the NGC 6822 
normal points, the isochrones were sampled at 0.5 mag increments in M$_K$, and 
the slopes measured from the results when M$_K \leq -3$ 
are summarized in Table 2, along with the [Fe/H] 
computed from Equation 1. The isochrones are steeper than the giant 
branches of actual clusters having the same metallicity, and this has been 
noted in other model sequences (e.g. Davidge 2000). The absolute calibration 
of the metallicity scale notwithstanding, the metallicities computed from the 
log(t) = 9.3 and 10.2 sequences differ by only 0.16 dex, whereas 
the metallicities computed from the log(t) = 9.6 and 10.2 isochrones differ by 
only 0.05 dex. Therefore, if RGB stars in NGC 6822 have a typical age of 
3 Gyr, then the metallicity from the RGB slope may be underestimated 
by $\sim 0.05$ dex, which is smaller than the random uncertainies in the 
metallicity estimate.

	The disk of NGC 6822 likely contains stars spanning a range of ages and 
metallicities, and constraints on possible age and metallicity dispersions can 
be estimated from the scatter in the composite CMD on the upper portion of the 
RGB. The RGB at M$_K = -5.5$ is well populated, and so the width of the CMD 
near this point is used to set constraints on any age and/or metallicity 
spread in the disk of NGC 6822. The standard deviation in $(J-K)_0$ of 
stars with M$_K$ between --5.4 and --5.6 is $\pm 0.07$ mag. The artificial 
star experiments indicate that photometric errors contribute 
a $\pm 0.03$ mag scatter in $J-K$ at this brightness. 
After subtracting this contribution in quadrature from the observed 
color dispersion there then remains $\pm 0.06$ mag scatter in $(J-K)_0$ that 
is not due to photometric errors. Comparisons with the F2000 
globular cluster sequences indicate that this range in color 
could result from a $\pm 0.2$ dex metallicity spread, while the 
G2000 isochrones indicate that a $\pm 0.5$ dex spread in 
log(t$_{Gyr}$) could also explain the observed scatter. The actual 
spreads in age and metallicity in the disk of NGC 6822 will of course be 
smaller than these values, as there is likely a dispersion in both age and 
metallicity. In addition, differential reddening and anisoplanicity may also 
contribute to the width of the RGB.

\section{DISCUSSION \& SUMMARY}

	The slope of the RGB on the $(K, J-K)$ CMD indicates that 
RGB stars in the disk of the Local Group dIrr galaxy NGC 6822 have a 
typical metallicity [Fe/H] $= -1.0 \pm 0.3$. This 
agrees with the mean metallicity measured for giants by Tolstoy 
et al. (2001) from the strength of the near-infrared Ca II 
absorption lines. The agreement between the metallicities determined from 
the RGB slope and the Ca II lines is noteworthy, as they are 
based on independent datasets that have different sensitivities to factors 
such as reddening and chemical mixture.

	Internal reddening is significant in some parts of NGC 6822. 
Multicolor observations of massive stars indicate that $E(B-V)$ varies between 
0.25 and 0.45 (Bianchi et al. 2001); for comparison, 
Wilson (1992) finds that $E(B-V) = 0.45$, which is much higher than the 
foreground value, based on the ridgeline of upper main 
sequence stars. Internal reddening is highest near star-forming regions, and 
studies of the colors of young stars with known 
spectral types suggest that there is a systematic reddening gradient in NGC 
6822, such that $E(B-V) = 0.45$ near the middle of the galaxy, which is the 
most active star-forming region, and $E(B-V) = 0.26$ at the 
periphery (Massey et al. 1995). This is qualitatively consistent with what 
was measured by Wyder (2001), who found that 
$E(B-V) = 0.34$ near the bar and $E(B-V) = 0.23$ in the disk. 
These studies thus indicate that foreground dust is the dominant source of 
reddening in the disk of NGC 6822.

	The RGB slope determined from the $(K, J-K)$ CMD is a relative 
measurement made from a self-contained dataset, and hence is not 
sensitive to the mean line-of-sight reddening, although differential reddening 
will smear the RGB, and complicate efforts to define the RGB 
ridgeline. The metallicity estimated from Ca II line strengths is 
moderately sensitive to the line-of-sight reddening. The relative $V$ 
brightness of RGB stars with respect to the horizontal branch (HB) is used 
to correct for surface gravity effects in Ca II line strengths 
(e.g. Armandroff \& Da Costa 1991) and, lacking deep photometry, 
Tolstoy et al. (2001) adopted a single HB brightness with only the 
foreground reddening value. The Tolstoy et al. (2001) field 
includes the bar, so some of the stars may be subject to reddening 
that exceeds the foreground value. By adopting a fixed HB brightness, 
Tolstoy et al. (2001) underestimated the metallicities of these stars. 
For RGB stars outside of active star-forming 
regions this effect will not be large; however, for stars with $E(B-V) = 0.45$, 
which appears to be the upper limit of line-of-sight reddening in NGC 6822, the 
metallicity computed by Tolstoy et al (2001) is be 0.16 dex too low. The 
main consequence is that the mean metallicity found by Tolstoy et al. (2001) is 
a lower limit, while the metallicity dispersion determined from their data 
($\pm 0.5$ dex, compared with $\pm 0.2$ dex from the width of the RGB in \S 4) 
is an upper limit.

	The use of the Ca II lines to measure [Fe/H] assumes that the [Ca/Fe] 
ratio matches that of the globular clusters that are used to define the 
metallicity calibration. This assumption may not hold, as the chemical 
enrichment history of galaxies that have experienced star formation, and 
hence chemical enrichment, over a large fraction of the age of the Universe, 
will liklely have been systematically different from that of globular clusters. 
For comparison, the overall photometric properties of the RGB, including 
the slope, are affected by the total metal content, rather than 
that of a single species.

	When applied to systems more distant than the Milky-Way and its 
immediate companions, metallicities based on the RGB slope and Ca II line 
strengths rely on observations of stars that are relatively faint, and hence 
may be susceptible to large photometric errors and systematic effects 
introduced by crowding. For example, the RGB must be traced down to the HB 
to measure the RGB slope. However, the fields studied here 
are in the low density outer regions of NGC 6822, and are not susceptible to 
crowding. In fact, each of the fields contain $200 - 
300$ stars on the upper 4 mag in $K$ of the RGB, so that the stellar density is 
$\sim 0.2$ stars arcsec$^{-2}$, or only 0.004 stars per angular resolution 
element, as defined by the FWHM of the PSF.

	The metallicity calibration used in the current paper assumes that RGB 
stars in NGC 6822 have ages comparable to those of globular clusters. In \S 4 
it is argued, based on the color of the RGB, that the majority of RGB stars in 
the NGC 6822 disk have an age of roughly 3 Gyr. While 
the model isochrones shown in the right hand panel of 
Figure 10 indicate that the slope of the RGB on the $(K, J-K)$ CMD is not 
sensitive to age at a fixed metallicity, if subsequent photometric 
measurements verify the relatively young ages of RGB stars in NGC 6822 then 
the metallicity derived from Equation 1 will be $\sim 0.05$ dex too low. 

	The giant branch in the composite NGC 6822 CMD of Fields 1, 2, and 3 is 
actually a mix of AGB and RGB stars, and the presence of AGB stars may affect 
the slope measurement and the color of the RGB locus. 
A distinct AGB sequence is not seen in the CMDs, because 
photometric errors blend the AGB and RGB. Nevertheless, the AGB has bluer 
colors than the RGB, with the color difference between the AGB and RGB 
decreasing towards the RGB-tip. Therefore, AGB stars may systematically bias 
the slope measurement and shift the RGB locus to bluer colors. While 
we have not attempted the difficult task of quantifying the effects of AGB 
contamination on the color and slope of the RGB, it should be noted that 
the number of AGB stars in a given brightness interval is 
much lower than that of RGB stars ($\frac{n_{AGB}}{n_{RGB}} \sim 0.25$), and 
the iterative rejection technique used to compute the normal points will 
reject objects with obvious outlying colors. 

	AGB stars could also cause the brightness of the RGB-tip to be 
over-estimated. This issue can be investigated 
by comparing the observed number of stars in a given brightness interval with 
those expected from a combined RGB$+$AGB population, which occurs below the 
RGB-tip, and a pure AGB population, which occurs above 
the RGB-tip. The dotted line in the lower panel of Figure 8 
shows the expected number counts from a pure AGB population. This sequence was 
created by shifting the power-law derived in \S 3.3 from faint RGB stars 
along the vertical axis to represent a
population that has 20\% of the total number of faint RGB $+$ AGB stars. 
When $K > 16.8$ (M$_K > -6.8$) each entry in the 
composite LF consistently falls above the AGB trend, as expected if these 
stars are a  mix of AGB and RGB stars, rather than only AGB stars. This 
comparison supports the adopted RGB-tip brightness of M$_K = -6.5$ that 
was measured in \S 4. 

	We close by noting that Muschielok et al. (1999) and Venn et al. (2001) 
conclude that [Fe/H] $= -0.5 \pm 0.2$ in bright NGC 6822 supergiants. These 
stars are considerably younger than RGB stars, for which [Fe/H] $= -1.0 \pm 
0.3$ based on the RGB slope. The present data thus 
suggest that RGB stars are $0.5 \pm 0.4$ dex more metal-poor 
than the youngest stars in NGC 6822. If the dominant RGB population has an age 
of 3 Gyr then these data indicate that $\frac{\Delta [Fe/H]}{\Delta t} = 
-0.2 \pm 0.1$ dex Gyr$^{-1}$ since intermediate epochs in the disk of 
NGC 6822.

\clearpage

\begin{center}
FIGURE CAPTIONS
\end{center}

\figcaption[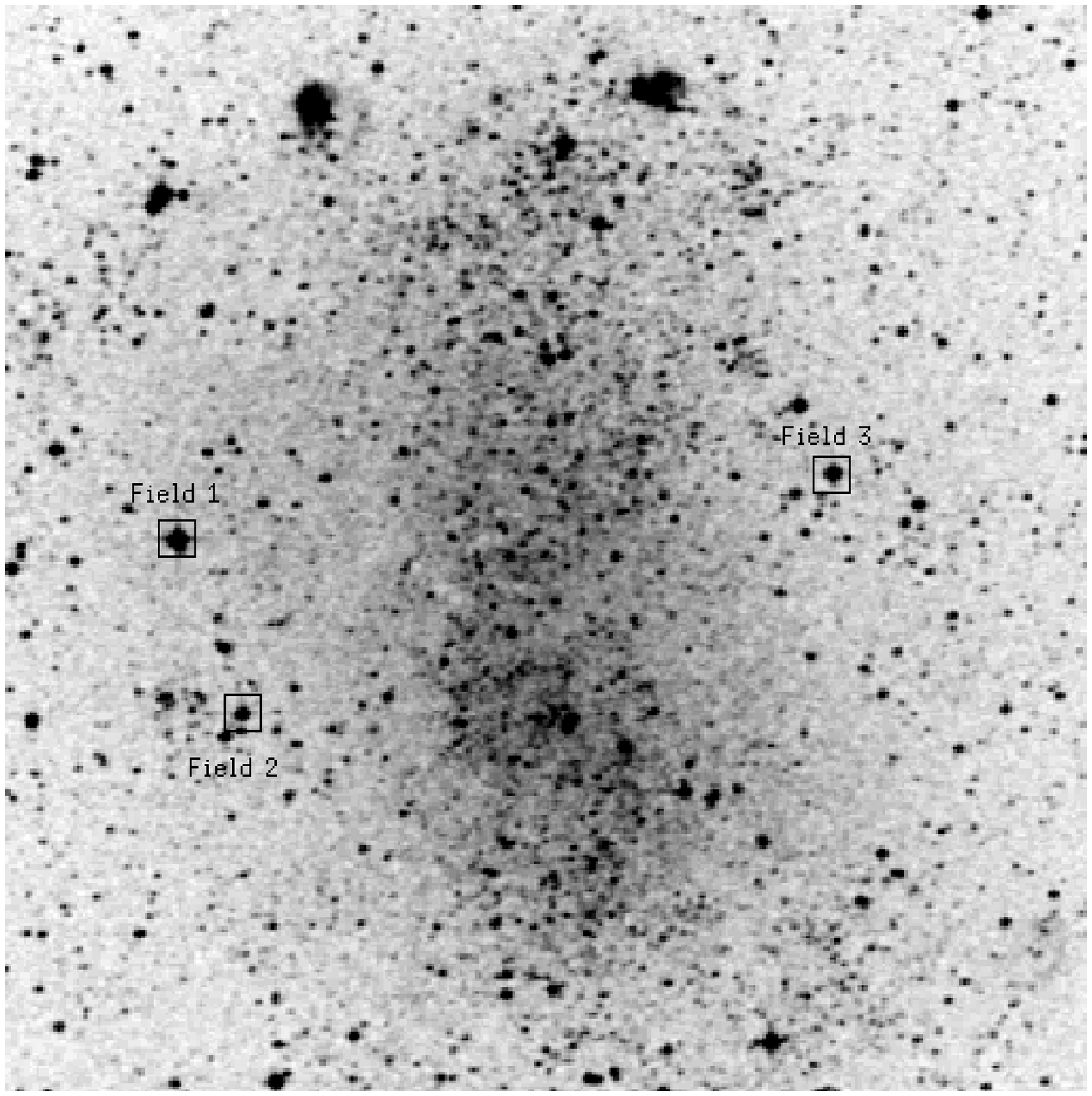]
{A $10 \times 10$ arcmin portion of the Digital Sky Survey 
centered on NGC 6822. The locations of the three fields, each of which is 
centered on a star that is bright enough to serve as a reference beacon for 
the AOB, are indicated. The boxes are approximately the same size as the KIR 
science field. North is at the top, and East is to the left.}

\figcaption[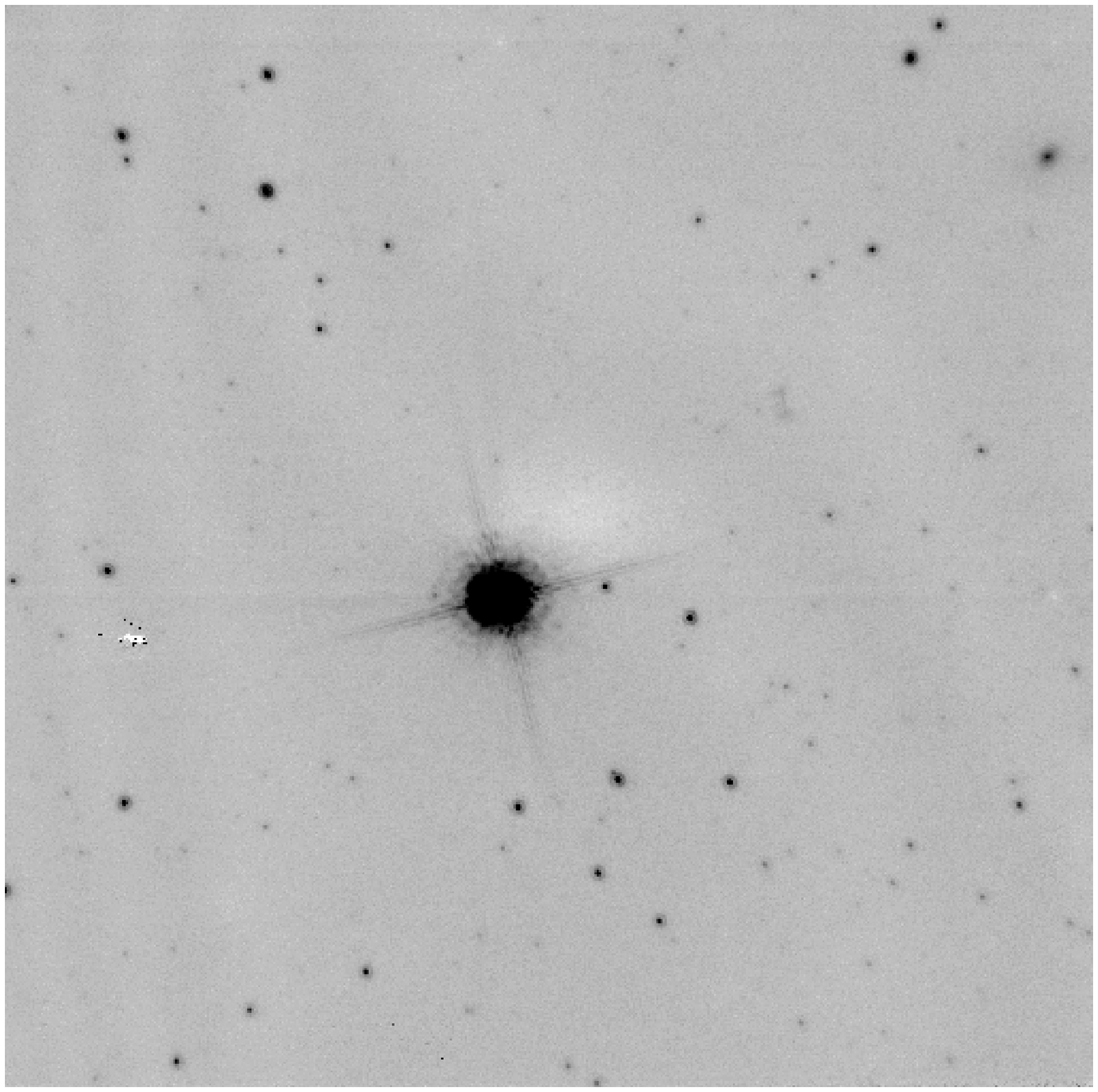]
{The final $K'$ image of Field 1. The image 
covers $34 \times 34$ arcsec, with North at the top, and East to the left. The 
bright central source is the AO guide star. Note that the first diffraction 
ring can be seen around the brightest stars.}

\figcaption[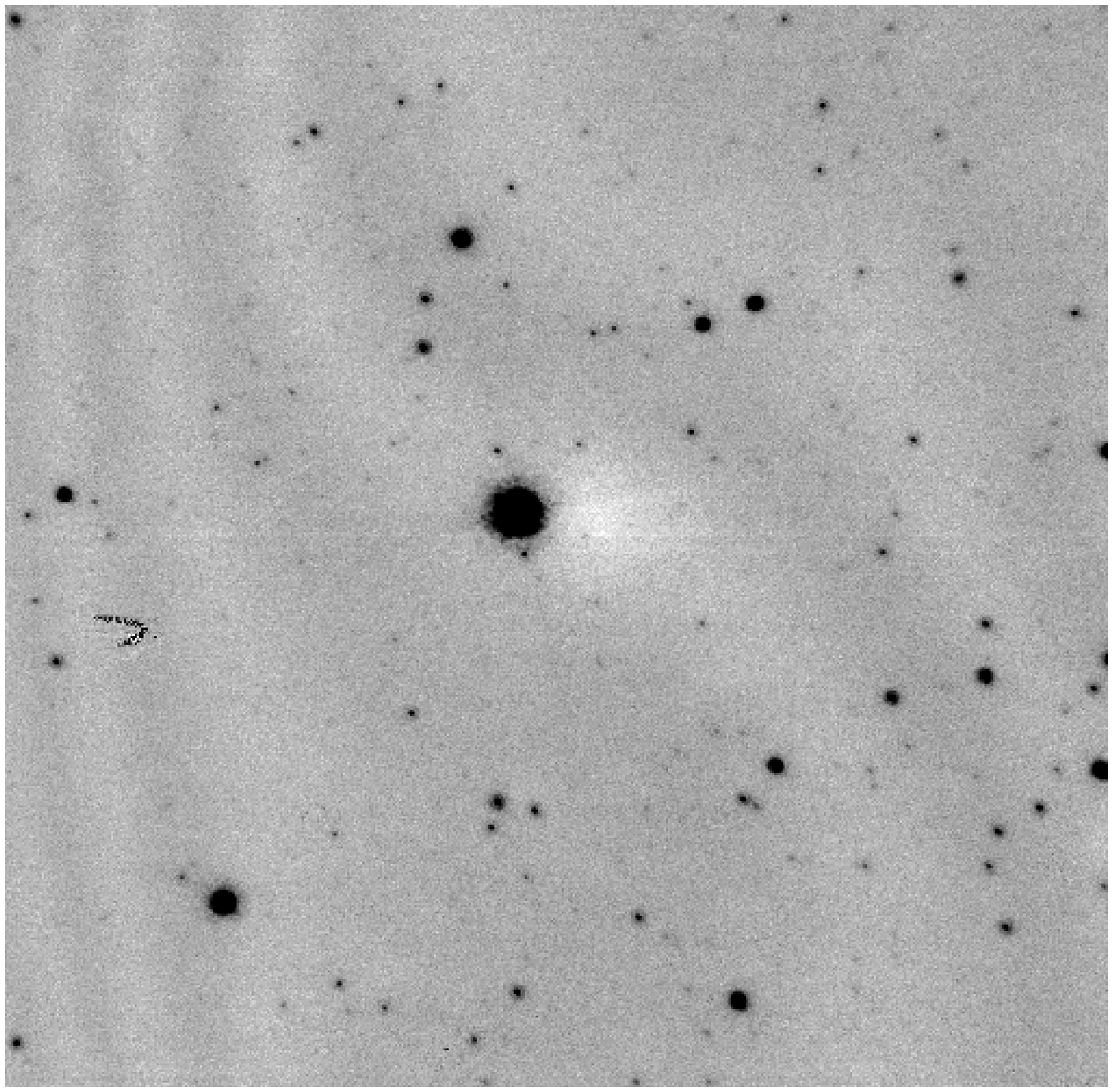]
{Same as Figure 2, but for Field 2.}

\figcaption[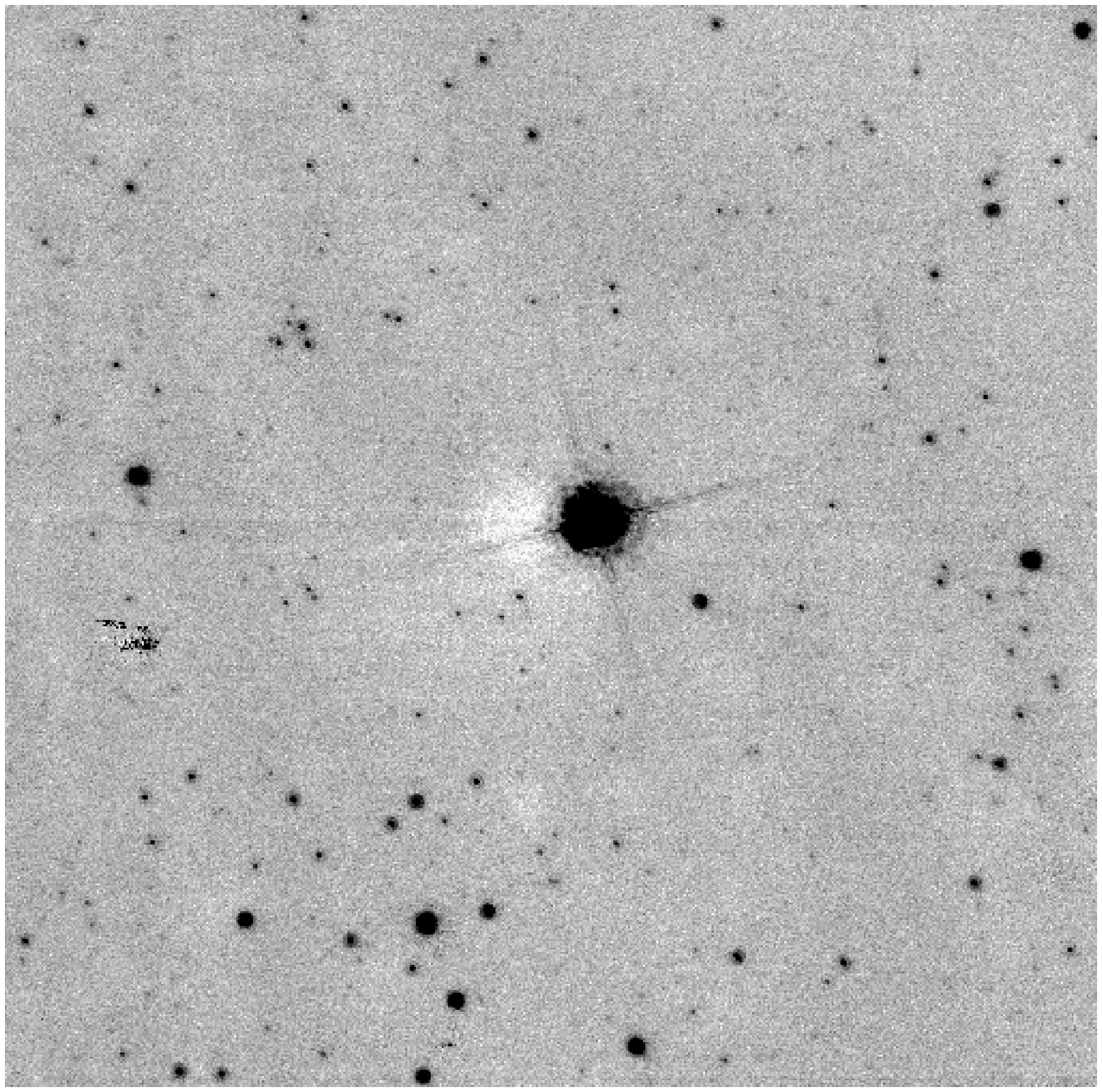]
{Same as Figure 2, but for Field 3.}

\figcaption[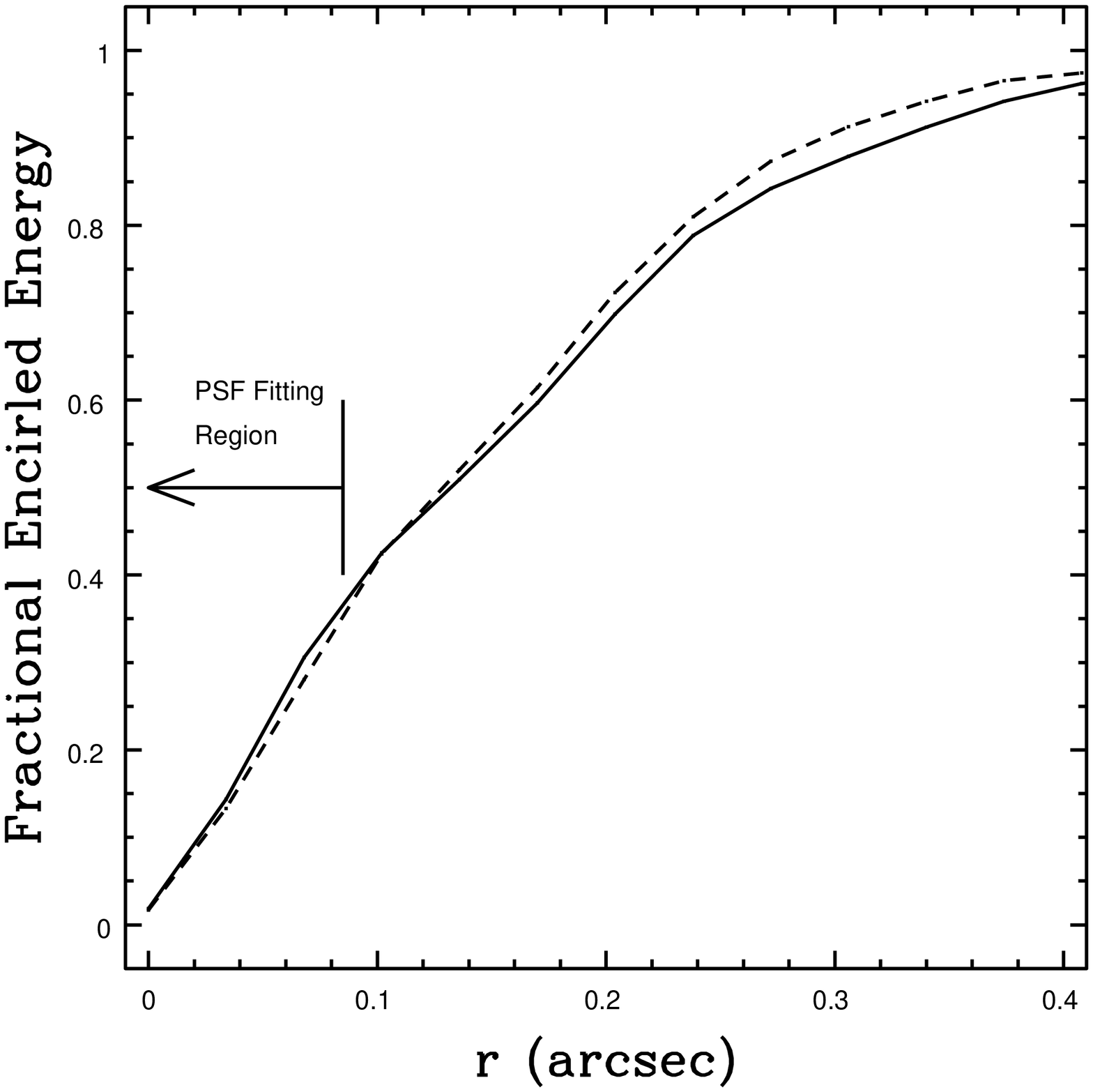]
{The encircled energy curves of PSFs that were 
constructed from stars within two different radial intervals in the Field 1 
$K'$ image. The solid curve shows the encircled energy curve, normalised 
to the total energy, for stars within 13.5 arcsec of the AO guide star, 
while the dashed line shows the same curve for stars at distances 
greater than 13.5 arcsec. The radial interval used for PSF-fitting 
is indicated, and the two curves agree to within a few percent in this 
region. This comparison indicates that anisoplanicity affects the photometry 
at only the few percent level.}

\figcaption[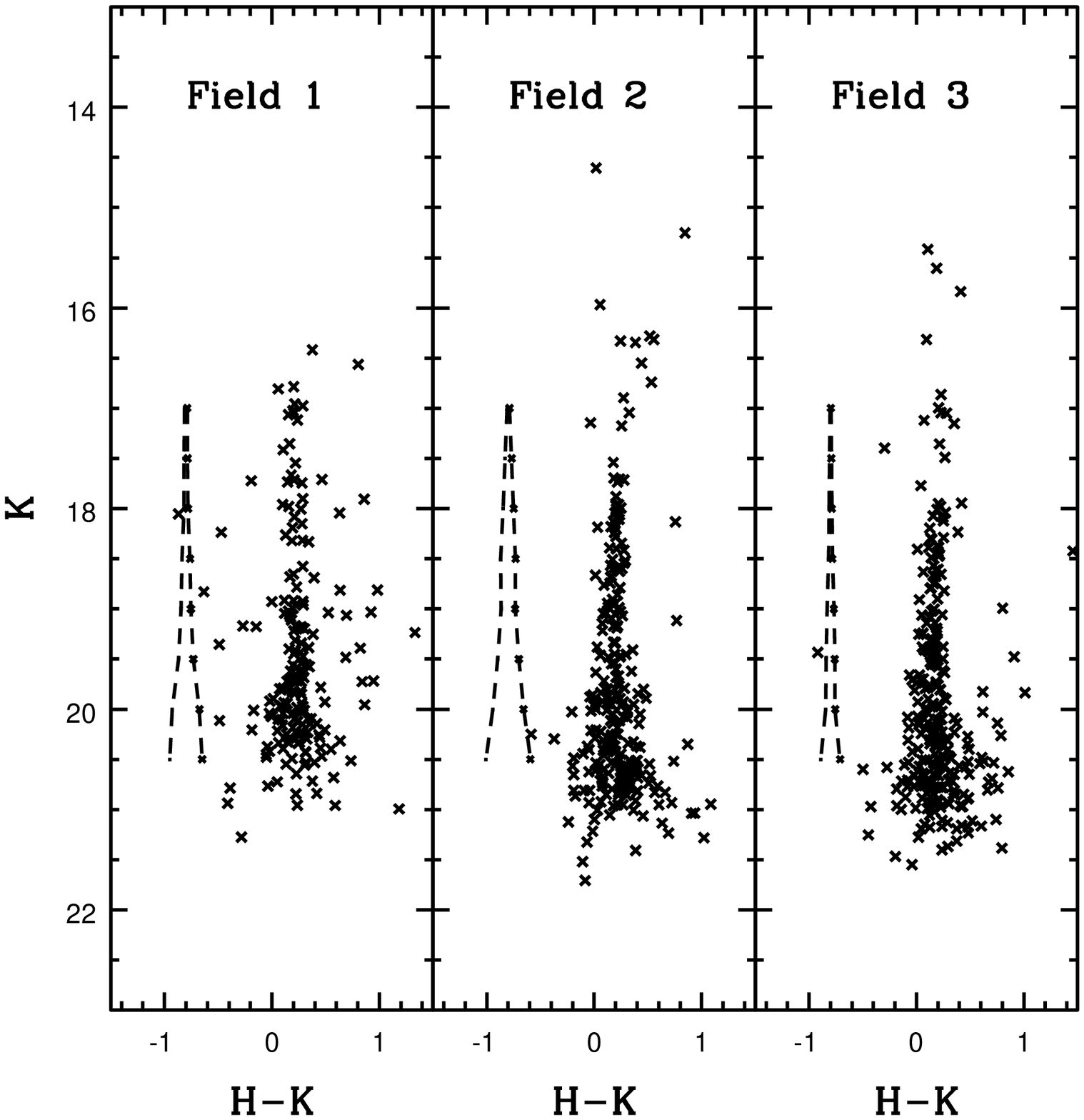]
{The $(K, H-K)$ CMDs of the three NGC 6822 fields. The RGB 
is the vertical sequence near $H-K = 0$ in each CMD. The dashed lines show the 
scatter envelope predicted from the artificial star experiments.} 

\figcaption[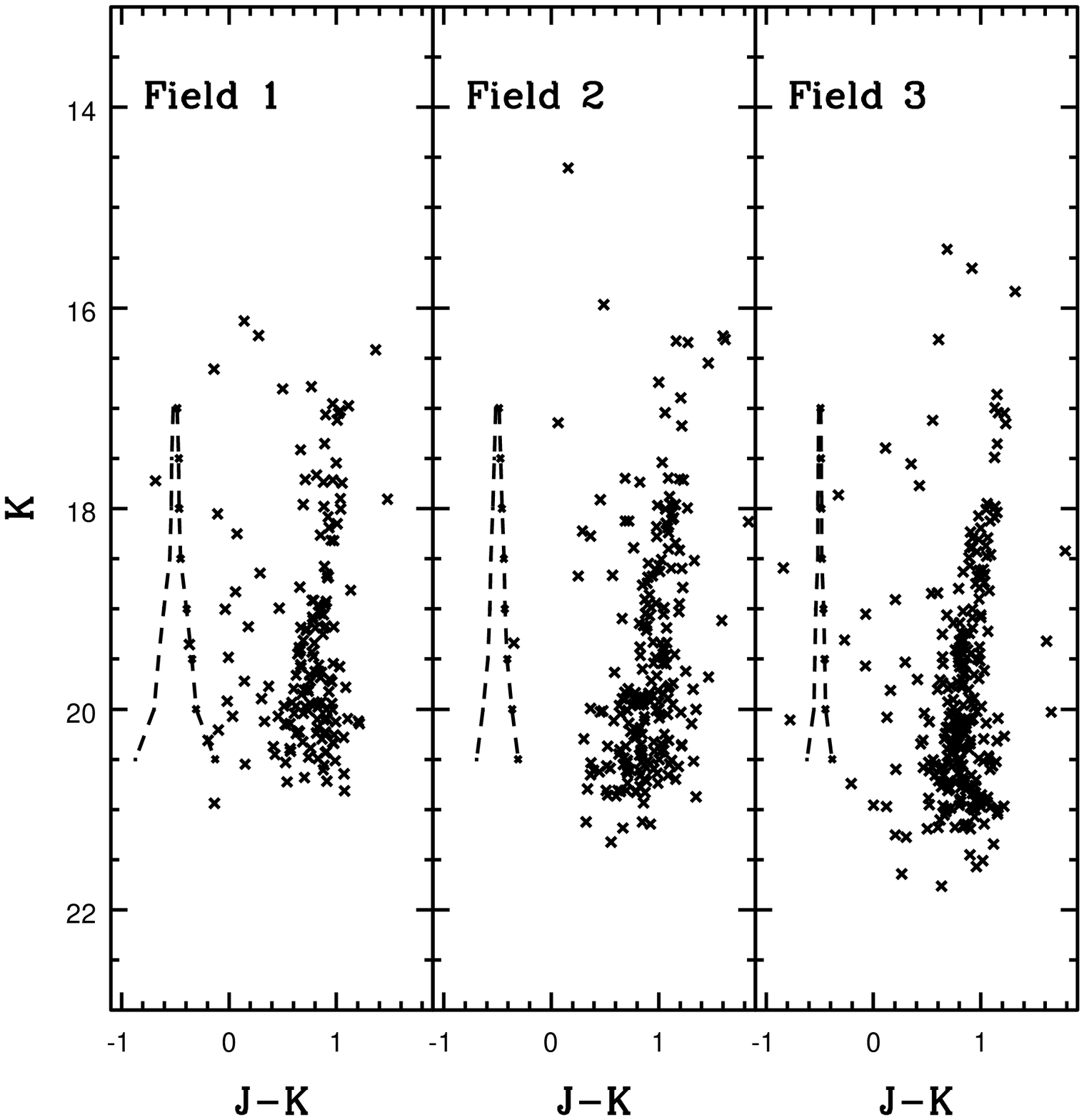]
{The $(K, J-K)$ CMDs of Fields 1, 2, and 3. The RGB 
is clearly seen in each CMD, and the RGB in all three fields terminates 
near $K = 17$. The dashed lines show the scatter envelope 
predicted from the artificial star experiments. While there is good agreement 
between the observed and predicted scatter at the faint end, when $K \leq 
19$ the observed scatter exceeds that predicted from photometric errors alone.}

\figcaption[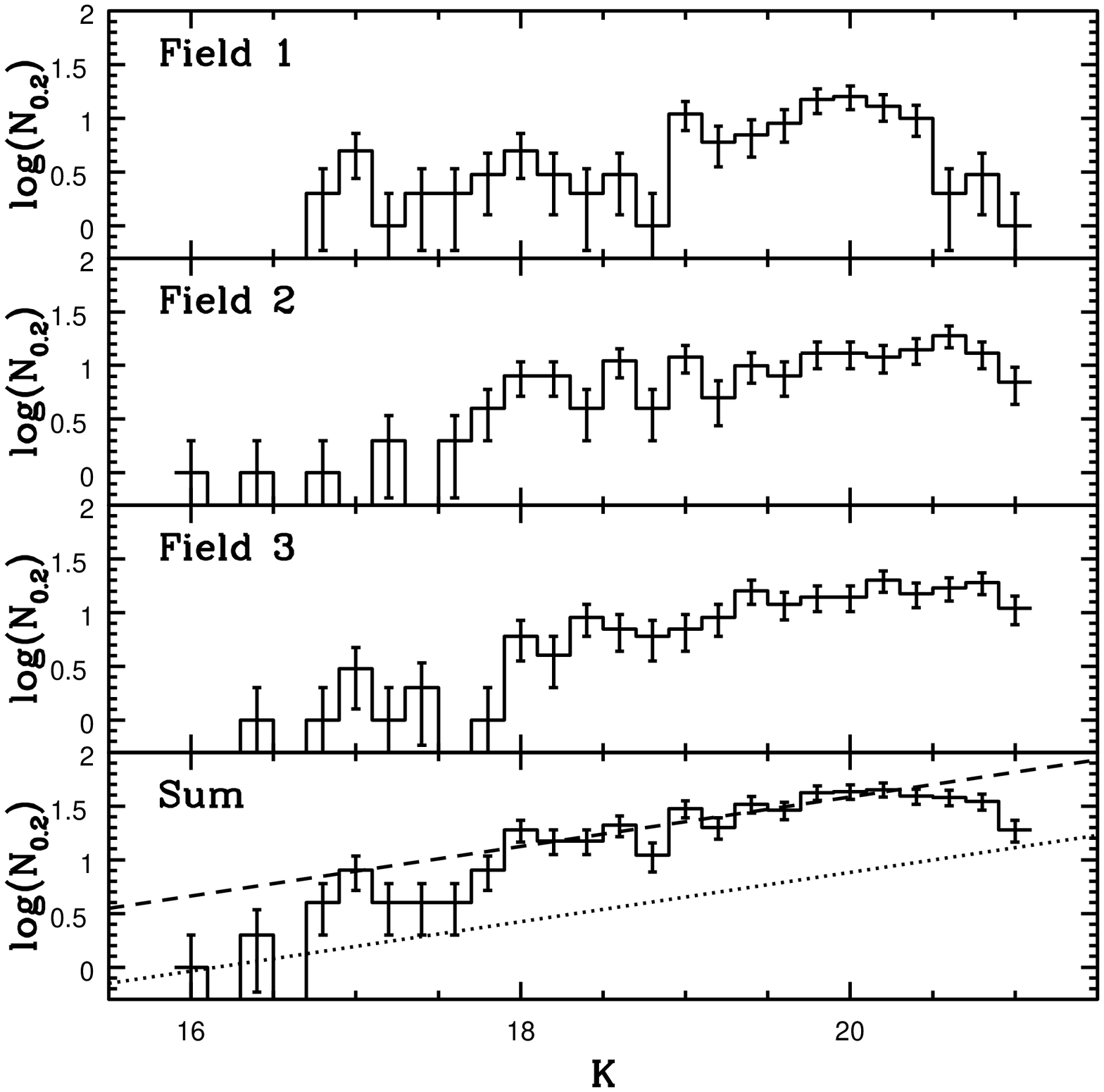]
{The $K$ LFs of Fields 1, 2, and 3, and their sum. N$_{0.2}$ is the number of 
stars per 0.2 magnitude interval in $K$. The dashed line in the bottom panel 
shows a power-law that was fit to the summed LF between $K = 18$ and 20; the 
power law has an exponent $0.23 \pm 0.06$. The dotted line shows this same 
power-law, but shifted to represent a population that is only 20\% the size 
of that on the observed LF, to simulate the contribution expected from a 
pure AGB component. Note that the observed LF consistently falls above 
the AGB sequence when $K > 16.8$ (M$_{K} > -6.8$), 
suggesting that this magnitude range is dominated by RGB stars.}

\figcaption[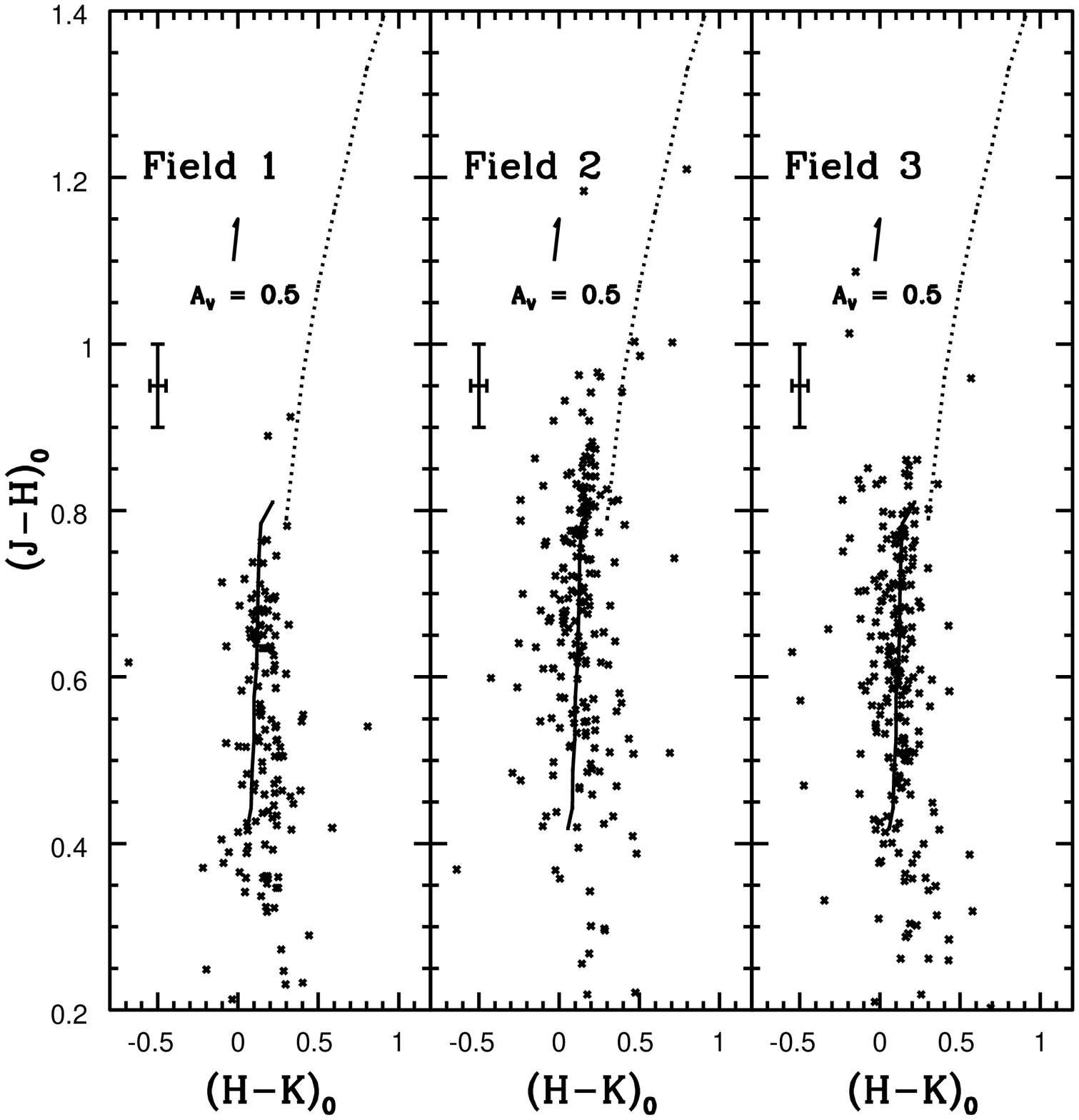]
{The $(J-H, H-K)$ TCDs of Field 1, 2, and 3. The data for each field have been 
corrected for foreground reddening, and the error bars show the 
uncertainty in the photometric calibration. A reddening vector with 
a length corresponding to A$_V = 0.5$ mag is also shown, and it is evident 
that extinction causes stars to move along an almost vertical track 
on this diagram. The solid line is the sequence for metal-poor globular cluster 
giants from Figure 12 of Davidge (2000), while the dotted line is the 
sequence defined for LMC LPVs from Figure 9 of Davidge (1998).} 

\figcaption[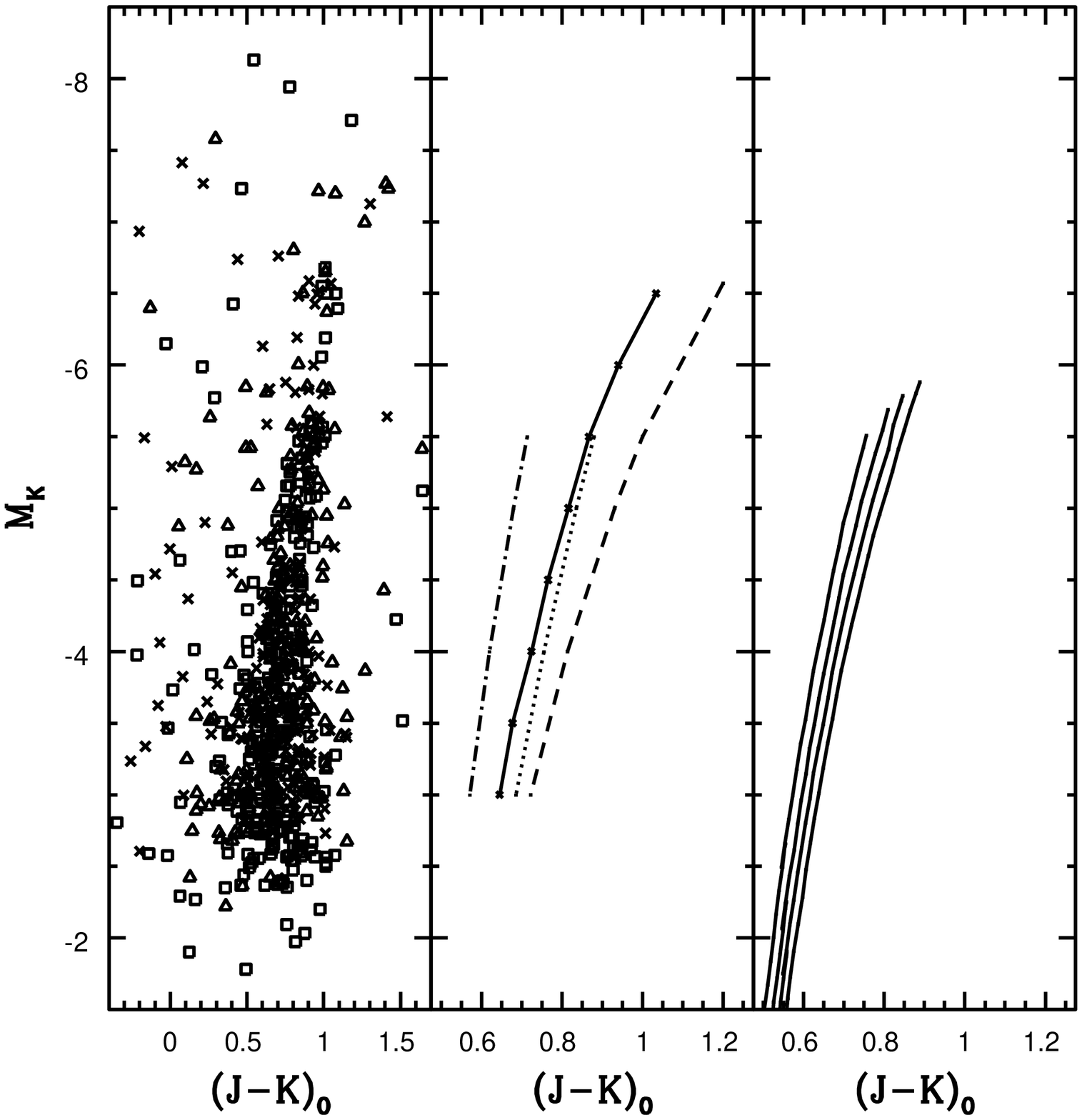]
{The left hand panel shows the composite $(M_K, (J-K)_0)$ 
CMD of Fields 1, 2, and 3. The data from Fields 1, 2, and 3 are plotted as 
crosses, triangles, and squares respectively. 
A foreground reddening of $E(B-V) = 0.24$ (Schlegel et 
al. 1998) and a distance modulus of 23.49 (Gallart 
et al. 1996) have been assumed. The solid line in the middle panel shows 
the normal point sequence calculated from the data in the left hand panel. 
Also shown in this panel are the RGB sequences for 47 Tuc (dashed line), 
NGC 6121 (dotted line), and NGC 7078 (dashed-dotted line) from F2000. Note that 
the NGC 6822 RGB falls blueward of the 47 Tuc and NGC 6121 sequences, even 
though the metallicity measured from the RGB slope falls between the two 
clusters. The left hand panel shows z=0.001 isochrones extending up to the 
RGB-tip for log(t$_{Gyr}$) = 9.3, 9.6, 9.9, and 10.2 from G2000. 
These models indicate that the difference in color between the RGB ridgeline in 
NGC 6822 and globular clusters having the same metallicity could occur if the 
majority of RGB stars in NGC 6822 have an age near 3 Gyr.}

\clearpage

\begin{table*}
\begin{center}
\begin{tabular}{lccc}
\tableline\tableline
Field \# & GSC$^{a}$ & RA & Dec \\
  & & (E2000) & (E2000) \\
\tableline
1 & 05736-00906 & 19:45:10.7 & --14:47:10 \\
2 & 05736-00192 & 19:45:08.1 & --14:48:46 \\
3 & 05736-00248 & 19:44:45.9 & --14:46:36 \\
\tableline
\end{tabular}
\end{center}
\caption{Field Co-ordinates}
\tablenotetext{a}{Guide Star Catalogue number of the AO reference star.}
\end{table*}

\clearpage

\begin{table*}
\begin{center}
\begin{tabular}{ccc}
\tableline\tableline
log(t$_{Gyr}$) & $\frac{\Delta (J-K)}{\Delta K}$ & [Fe/H]$^{b}$ \\
\tableline
9.3 & $-0.074 \pm 0.002$ & --1.62 \\
9.6 & $-0.080 \pm 0.004$ & --1.51 \\
9.9 & $-0.083 \pm 0.004$ & --1.46 \\
10.2 & $-0.083 \pm 0.004$ & --1.46 \\
\tableline
\end{tabular}
\end{center}
\caption{Slopes and [Fe/H] computed from the z = 0.001 isochrones of G2000.}
\tablenotetext{b}{Computed from Equation 1.}
\end{table*}

\end{document}